# ATOMS: ALMA Three-millimeter Observations of massive Star-forming regions -XX. Probability distribution function of integrated intensity for dense molecular gas tracers

C. Zhang,[1,2]* Tie Liu,[3]† Sihan Jiao,[4] Feng-Yao Zhu,[5] Z.-Y. Ren, [4] H.-L. Liu,[6] Ke Wang,[7] J.-W. Wu,[5,9]
D. Li,[8,9] P. García,[10,11] Guido Garay,[12,10] Leonardo Bronfman,[12] Mika Juvela,[13] Swagat das,[12]
Chang Won Lee,[14,15] Feng-Wei Xu,[16,17,18] L. V. Tóth,[19,20] Prasanta Gorai,[21] Patricio Sanhueza,[22,23]

[1] *Department of Physics, Taiyuan Normal University, Jinzhong 030619, China*
[2] *Institute of Computational and Applied Physics, Taiyuan Normal University, Jinzhong 030619, China*
[3] *Shanghai Astronomical Observatory, Chinese Academy of Sciences, 80 Nandan Road, Shanghai 200030, China*
[4] *National Astronomical Observatories, Chinese Academy of Sciences, Beijing, 100012, People's Republic of China*
[5] *Research Center for Astronomical Computing, Zhejiang Laboratory, Hangzhou, 311100, China*
[6] *Department of Astronomy, Yunnan University, Kunming,650091, PR China*
[7] *Kavli Institute for Astronomy and Astrophysics, Peking University, 5 Yiheyuan Road, Haidian District, Beijing 100871, People's Republic of China*
[8] *Department of Astronomy, Tsinghua University, Beijing 100084, China*
[9] *National Astonomical Observatories, Chinese Academy of Sciences, A20 Datun Road, Chaoyang District, Beijing 100101,China*
[10] *Chinese Academy of Sciences South America Center for Astronomy, National Astronomical Observatories, CAS, Beijing 100101, China*
[11] *Instituto de Astronomía, Universidad Católica del Norte, Av. Angamos 0610, Antofagasta, Chile*
[12] *Departamento de Astronom ́ıa, Universidad de Chile, Casilla 36-D, 8320168, Santiago, Chile*
[13] *Department of Physics, University of Helsinki, PO Box 64, FI-00014 Helsinki, Finland*
[14] *Korea Astronomy and Space Science Institute, 776 Daedeokdaero, Yuseong-gu, Daejeon 34055, Republic of Korea*
[15] *University of Science and Technology, Korea (UST), 217 Gajeong-ro, Yuseong-gu, Daejeon 34113, Republic of Korea*
[16] *Physikalisches Institut, Universität zu Köln, Zülpicher Str. 77, D-50937 Köln, Germany*
[17] *Kavli Institute for Astronomy and Astrophysics, Peking University, Beijing 100871, People's Republic of China*
[18] *Department of Astronomy, School of Physics, Peking University, Beijing, 100871, People's Republic of China*
[19] *Institute of Physics and Astronomy, Eotvos Lorand University, Pazmany Peter setany 1/A, H-1117 Budapest, Hungary*
[20] *University of Debrecen, Institute of Physics, H-4032 Debrecen, Bem tér 1*
[21] *Institute of Theoretical Astrophysics University of Oslo Sem Selands vei 13, 0371 Oslo*
[22] *Department of Earth and Planetary Sciences, Institute of Science Tokyo, Meguro, Tokyo, 152-8551, Japan*
[23] *National Astronomical Observatory of Japan, National Institutes of Natural Sciences, 2-21-1 Osawa, Mitaka, Tokyo 181-8588, Japan*

**ABSTRACT**

We report the observations of J=1-0 of HCN, HCO$^+$, H$^{13}$CO$^+$, and H$^{13}$CN, HC$_3$N (J=11-10) emission towards 135 massive star-forming clumps, as part of the ATOMS (ALMA Three-millimeter Observations of Massive Star-forming regions) Survey. We present the integrated intensity probability distribution function for these molecular tracers, modeled as a combination of a log-normal distribution and a power-law tail. The molecular line luminosities for the power-law tail segment, L$_{mol}(p)$, have been calculated. We have investigated the correlation between the bolometric luminosity, L$_{bol}$, and the power-law part of the molecular line luminosity, L$_{mol}(p)$. Our findings suggest that the scaling relationships between L$_{bol}$ and L$_{mol}(p)$ for HCN and HCO$^+$ are sublinear, indicating that these molecules might not be the most effective tracers for the dense gas. In contrast, H$^{13}$CN and HC$_3$N exhibit a nearly linear relationship between L$_{bol}$ and L$_{mol}(p)$, indicating that they can well trace gravitationally bound dense gas. The ratios of L$_{bol}$-to-L$_{mol}(p)$, serving as indicators of star formation efficiency within massive star-forming clumps, exhibit a weak anti-correlation with the power-law index in the I-PDF. In addition, the star formation efficiency is also weakly anti-correlated with the exponent $\alpha$ of the corresponding equivalent density distribution. Our results implie that clumps with substantial gas accumulation may still display low star formation efficiencies.

## 1 INTRODUCTION

Stars are known to form in molecular gas, but recent works suggest that only the dense gas should be responsible for star formation rate (SFR) (Evans 2008; Vutisalchavakul et al. 2016). On the scale of galaxies, Gao & Solomon (2004) showed that IR luminosities ($L_{IR}$) tracing the SFR on galactic scales have a tighter linear correlation with HCN (J=1-0) luminosities than with CO luminosities. This correlation extends toward smaller spatial scales, to Galactic dense clumps undergoing high-mass star formation (Wu et al. 2005). Subsequent studies have defined a "threshold" surface density of 120 $M_\odot$ pc$^{-2}$ in nearby clouds (Lada et al. 2012), above which the vast majority of dense cores and Young Stellar Objects (YSOs) are found. This so called "dense gas star formation law" suggests that it is dense molecular gas rather than the total molecular gas that directly fuels star formation. Linear correlations have been identified between the luminosities from infrared emissions and those from other dense gas tracers, including HCN (J=4-3), HCO$^+$ (J=4-3), CS (J=7-6), CS (J=2-1), H$^{13}$CO$^+$ (J=1-0), H$^{13}$CN (J=1-0), HC$_3$N (J=11-10)(Wu et al. 2010; Zhang et al. 2014; Liu et al. 2016, 2020b; Zhang et al. 2022).

* E-mail: zhangchao920610@126.com
† E-mail: liutie@shao.ac.cn

Although the tight connection between dense parts of molecular clouds and star formation is clear, the origin of this relationship is still under debate. The molecular line emission (e.g. HCN and HCO$^+$) are among the most commonly used tracers in the studies of the "dense gas star formation law". However, the molecular emission line, in addition to surface and volume densities, is sensitive to temperature and molecular abundances, which in turn depend on the radiation environment (Pety et al. 2017; Shimajiri et al. 2017). Moreover, Stephens et al. (2016) have argued that most of the Galaxy's luminosity of HCN may arise from extended, subthermal emission rather than from dense gas. More recently, Evans et al. (2020a) found that a substantial fraction (in most cases, the majority) of the total HCN (1-0) and HCO$^+$ (1-0) line luminosity in six distant ($d = 3.5$–10.4 kpc) clouds arises below the empirical "threshold" surface density (e.g., $A_V = 8$ mag; Lada et al. 2010; Evans et al. 2014). These studies challenge the idea that these commonly used dense-gas tracers can reveal the spatial distribution of the star-forming gas in molecular clouds.

A further issue with the "dense gas star formation law" is its vague definition of how dense the gas need to be to form stars. To be physically meaningful, the "dense gas" should be quantified more specifically, such as through the power-law tail of probability distribution functions (PDFs) of gas densities. PDFs form the basis of many modern theories of star formation (e.g. Krumholz & McKee 2005; Pan et al. 2019). Supersonic turbulent flows in isothermal, non-gravitating medium lead to a lognormal PDF. In the high-density range, the power-law tail out of a quasi-lognormal PDF reflects the increasing role of self-gravity in evolving star-forming clouds (Klessen 2000; Collins et al. 2012; Burkhart et al. 2017). Jaupart & Chabrier (2020) showed that the power-law tail density range corresponds to regions in which gravity is the main driver of turbulence statistics. Kainulainen et al. (2014) suggested that the probability density functions (PDFs) of volume densities initiate a density threshold for star formation and enable the quantification of star formation efficiency above this threshold. Additionally, similar approaches using the power-law tails of PDFs of column densities to quantify "dense gas fraction" were employed in other studies (Kainulainen et al. 2013; Schneider et al. 2013; Burkhart et al. 2017; Jiao et al. 2022). In a word, exploring the potential relationship between the gas within the power-law tail and the star formation process is crucial.

In this study, we aim to investigate the correlation between gravitationally bound gas and star formation by analyzing the integrated intensity probability distribution functions (I-PDFs) and the power-law tails within the luminosity distribution of molecular transitions such as HCN, HCO$^+$, H$^{13}$CO$^+$, H$^{13}$CN with J=1-0, CS with J=2-1, and HC$_3$N with J=11-10 across a sample of molecular clumps throughout the Galactic plane observed as part of the ATOMS Survey(Liu et al. 2020a).

## 2 OBSERVATIONS

The ATOMS survey, standing for ALMA Three-millimeter Observations of Massive Star-forming regions, have observed 146 active Galactic star-forming regions (Liu et al. 2020a). These 146 sources were selected from the CS (2-1) survey of IRAS UCH$_{II}$ region candidates with bright CS emission $T_b > 2$ K (Bronfman et al. 1996) and solid HCN follow-up detections (Liu et al. 2016). The global dust temperature, mass, and luminosity of these 146 targets were obtained from the infared spectral energy distribution (SED) fitting (Faúndez et al. 2004; Urquhart et al. 2018), and are categorized in Table A1 of Liu et al. (2020a). There are 139 targets located in the first and fourth Galactic Quadrants of the inner Galactic Plane. The rest are distributed at the outer Galaxy. The distances of the sample clouds range from 0.4 kpc to 13.0 kpc with a mean value of 4.5 kpc. The sample includes 27 distant ($d > 7$ kpc) sources that are either close to the Galactic Center like SgrB2(M) or mini-starbursts like W49A, representing extreme environments for star formation.

The ALMA data calibration and imaging were processed using the CASA (McMullin et al. 2007). The 12-m and ACA 7-m array data were jointly imaged using the Briggs weighting and robust of 0.5, in the CASA tclean task, for both continuum images and line cubes. For each source, eight data cubes are obtained, including six narrow bands and two wide bands to cover target lines (refer to Table 1 in Liu et al. 2020a). All images are primary-beam corrected. The typical beam sizes and maximum recovering scales (MRS) in the combined data are $2''$ and $1'$, respectively. The observations are not sensitive to extended structures larger than $1'$, therefore, diffuse emission in lines is mostly filtered out. In other words, the ALMA observations are more sensitive to dense and compact structures, which are more likely gravitationally bound.

We have identified 135 sources that exhibit emissions in six molecular transitions: HCN, HCO$^+$, H$^{13}$CO$^+$, and H$^{13}$CN with J=1-0, CS (J=2-1), HC$_3$N (J=11-10). To put these massive clumps into the similar physical resolution and discuss the characteristics of the power-law tails, the whole sample have been categorized into three distinct distance intervals. The intervals range from less than 3 kpc, between 3 and 6 kpc, and greater than 6 kpc. Within each distance interval, the sample sources have undergone a smoothing process to achieve a uniform linear resolution. The smoothing process was performed using the CASA version. The linear resolutions attained for the sources in intervals of less than 3 kpc, 3 to 6 kpc, and greater than 6 kpc are approximately 0.03 kpc, 0.07 kpc, and 0.15 kpc, respectively. Figure 1 illustrates the integrated intensity maps of molecular transitions for the source I11332-6258. We have generated images for all sources; the one displayed here is a sample, with the remainder to be made available online. From these figures, we can see that the emission of molecular lines is mostly concentrated within regions smaller than the MRS ($\sim 1'$), indicating that the missing flux is not an issue in investigating the spatial distribution of gravitationally bound gas within these clumps.

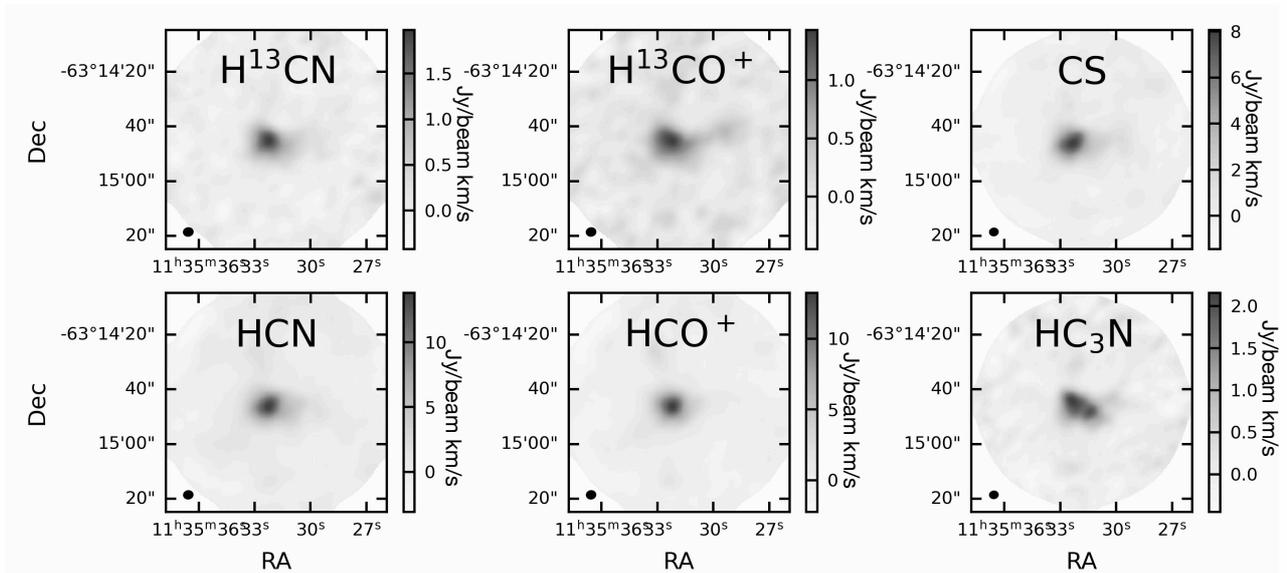

**Figure 1.** Integrated intensity images of emission at HCN, HCO$^+$, H$^{13}$CO$^+$, and H$^{13}$CN with J=1-0, CS (J=2-1), HC$_3$N (J=11-10) of source I11332-6258. The ellipse located at the lower left corner of each panel represents the beam size.

## 3 RESULTS

### 3.1 Integrated intensity probability distribution functions

We define the functional form of the integrated intensity probability distribution functions (I-PDFs) as:

$$PDF(I) = A_m exp[-(I - \mu)^2/(2\sigma^2)],$$

where $I$ is the integrated intensity at each pixel measured in units of Jy/beam km/s. $A_m$ is the amplitude, $\sigma$ is the dispersion of the logarithmic field, and $\mu$ is the mean value of the distribution. The parameter $\sigma$ is determined by fitting the assumed lognormal distribution to the low-intensity region of the I-PDF, while the slopes of the high-density regions are derived from a linear regression power-law fit given by $P[log(I)] = s * log(I) + c$. Hereafter, the term P[log(I)] denotes the function describing the power-law tail of the distribution.

Having fitted the low-intensity side of the peak with a log-normal distribution, we subsequently subtract this best-fit log-normal profile from the overall I-PDF. The remaining secondary component is well characterized by a power-law distribution. Figure 2 illustrates the PDFs of intensity for the six dense tracers of the exemplar source I11332-6258. All traces demonstrate a log-normal PDF and exhibit power-law tails.

Figure 3 displays the distribution of the power-law tail slopes $s$ for six tracers of all sources. Their values range approximately from -4 to -1.5. The median values of $s$ for H$^{13}$CN, H$^{13}$CO$^+$, CS, HCN, HCO$^+$, HC$_3$N are -1.98, -2.28, -2.01, -2.19, -2.10, -2.02, respectively. Table 1 presents the parameters for H$^{13}$CO$^+$, including $\sigma$ and $\mu$ and $s$, across columns two to four. Parameters for all tracers including $\sigma$ and $\mu$ and $s$, along with the full list of sources, can be accessed online.

### 3.2 The L$_{bol}$-L$_{mol}$ scaling relations

The bolometric luminosity (L$_{bol}$) has been widely used as a tracer of the recent SFR (Gao & Solomon 2004; Wu et al. 2005, 2010). L$_{bol}$ values for the source in this work are calculated by integrating all spectral energy distributions (Faúndez et al. 2004; Urquhart et al. 2018). Table 1 presents L$_{bol}$ in the seventh column. Concurrently, the line luminosities (L$_{mol}$) of dense molecular gas are tracers of the dense molecular gas mass in star-forming regions or external galaxies (Gao & Solomon 2004; Wu et al. 2005). These luminosities are calculated from the observed line fluxes and are derived as below (Zhang et al. 2022):

$$L_{mol} = 32.5 \nu_{obs}^{-2} S_{mol} \Delta v D^2,$$  (1)

where $L_{mol}$ is the the molecular luminosity in K km/s pc$^2$, $S_{mol}\Delta V$ is the velocity-integrated flux in Jy km s$^{-1}$, $\nu_{obs}$ is the line frequency in GHz, $D$ is the distance in kpc. In this work, we focus on the dense gas traced by the power-law tails in the I-PDFs, which is more likely gravitationally bound gas. The line luminosity attributed to the power-law high-intensity component is termed L$_{mol}$(p), to be distinguished from the total line luminosity used in previous works (e.g., Liu et al. 2020b; Zhang et al. 2022). L$_{mol}$(p) for H$^{13}$CO$^+$ J=1-0 are presented in the eighth columns of Table 1. The comprehensive line luminosity data for additional tracers along with all sources are accessible online.

Previous studies (e.g., Wu et al. 2005, 2010; Liu et al. 2016, 2020b; Zhang et al. 2022) adopted the total line luminosities to investigate the dense gas star formation scaling relations. In contrast, we use the line luminosity of the power-law tail component in I-PDF, i.e., L$_{mol}$ (p). Figure 5 shows the scaling relations L$_{bol}$-to-L$_{mol}$ (p) of six tracers. All relations are determined by linear least-squares fits in log–log space. Relations with slopes near unity and within the range of twice the error margin in logarithmic fits are referred to as linear, while those with slopes significantly less than unity are referred to as sublinear. The fit towards the six different tracers is shown in black dash line. The slope of the correlation for H$^{13}$CN, H$^{13}$CO$^+$, CS, HCN, HCO$^+$, HC$_3$N are 1.0 ± 0.07, 0.94 ± 0.07, 0.93 ± 0.07, 0.82 ± 0.09, 0.83 ± 0.08, 0.98 ± 0.08, respectively. The green dashed line represents a one-to-one correspondence. It shows that the regression slopes of the fits for HCN and HCO$^+$ exhibit a distinct sublinear trend. Whereas the other tracers indicate that the relations are close to linear.

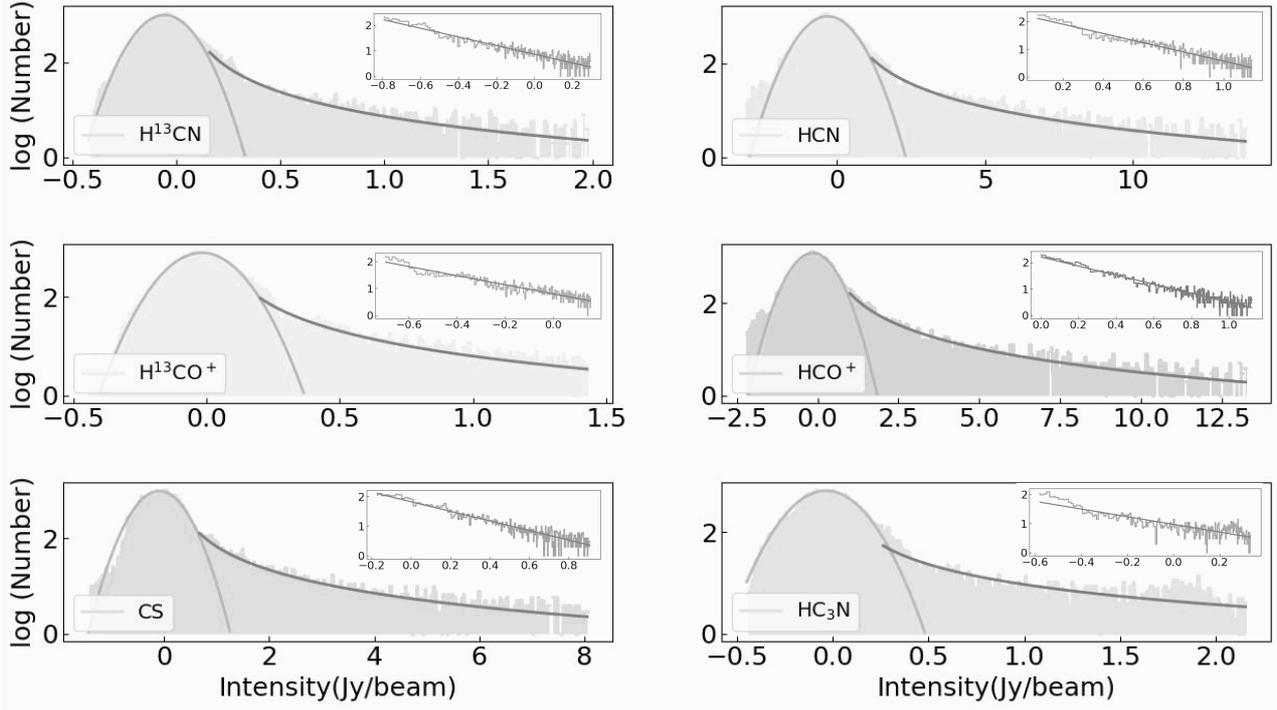

**Figure 2.** I-PDFs of emission at J=1-0 of HCN, HCO$^+$, H$^{13}$CO$^+$ and H$^{13}$CN, and CS J=2-1, HC$_3$N J=11-10 for source I11332-6258. The y-axis is presented on a logarithmic scale to enhance the visualization of features within the lognormal distribution. Each subplot includes an inset in the upper left corner, illustrating the power-law tail with the abscissa on a logarithmic scale to accentuate the characteristics of the high-intensity domain.

| Source Name | $\mu$ | $\sigma$ | s | distance (kpc) | log(M$_{clump}$) M$_\odot$ | log(L$_{bol}$) L$_\odot$ | log(L$_{mol}$(p)) (K km/s pc$^2$) |
|---|---|---|---|---|---|---|---|
| I08076-3556 | 0.281 ± 0.054 | 1.369 ± 0.057 | -0.75 ± 0.045 | 0.4 | 0.7 | 1.20 | -2.57 |
| I08303-4303 | -0.005 ± 0.003 | -0.187 ± 0.003 | -3.35 ± 0.096 | 2.3 | 2.4 | 3.83 | -0.67 |
| I08470-4243 | -0.011 ± 0.003 | -0.178 ± 0.003 | -5.454 ± 0.162 | 2.1 | 2.4 | 4.04 | -1.09 |
| I09018-4816 | -0.014 ± 0.002 | -0.202 ± 0.002 | -3.317 ± 0.1 | 2.6 | 3.0 | 4.72 | -0.56 |
| I09094-4803 | -0.002 ± 0.001 | 0.112 ± 0.001 | -5.094 ± 0.262 | 9.6 | 3.1 | 4.60 | -0.38 |

**Table 1.** The paramenters of H$^{13}$CO$^+$ J=1-0. The full catalogue is available on line.

Are the different regression slopes caused by optical depths? To investigate this issue, we have compared the optical depths of HCO$^+$ and H$^{13}$CO$^+$, as well as HCN and H$^{13}$CN. The brightness temperature T$_b$ of a molecular line transition is:

$$T_b = f[J_\nu(T_{ex}) - J_\nu(T_{bg})](1 - e^{-\tau}) \tag{2}$$

where $J_\nu(T) = \frac{h\nu}{k_B} \frac{1}{exp(h\nu/kT)-1}$. $f$ is the fraction of the telescope beam filled by the emission, or beam-filling factor. $f$ is taken to be

1 since ALMA can well resolve the gas structures (Xu et al. 2023). $T_{bg}$ = 2.73 K is the temperature of the cosmic background radiation.

Assuming local thermodynamic equilibrium (LTE) and the same excitation temperature for HCO$^+$ and H$^{13}$CO$^+$, the optical depth of HCO$^+$ can be directly determined by comparing the observed brightness temperatures of HCO$^+$ and H$^{13}$CO$^+$ (Garden et al. 1991):

$$\frac{T_{b,HCO^+(\nu)}}{T_{b,HCO^+(\nu)}} = \frac{1 - exp(-\tau_{\nu,12})}{1 - exp(-\tau_{\nu,13})} = \frac{1 - exp(-\tau_{\nu,12})}{1 - exp(-\tau_{\nu,12}/(X_{12}/X_{13}))} \tag{3}$$

where $\tau_{\nu,12}$ and $\tau_{\nu,13}$ are the optical depths of the HCO$^+$ and H$^{13}$CO$^+$ J=1-0, respectively. We assumed the abundance ratio X$_{12}$/X$_{13}$=[HCO$^+$]/[H$^{13}$CO$^+$] as a function of Galactocentric distance as given by Pineda et al. (2013),

$$\frac{X_{12}}{X_{13}} = 4.7\left(\frac{R_{gal}}{kpc}\right) + 25.05 \tag{4}$$

where $R_{gal}$ is the Galactocentric distance.

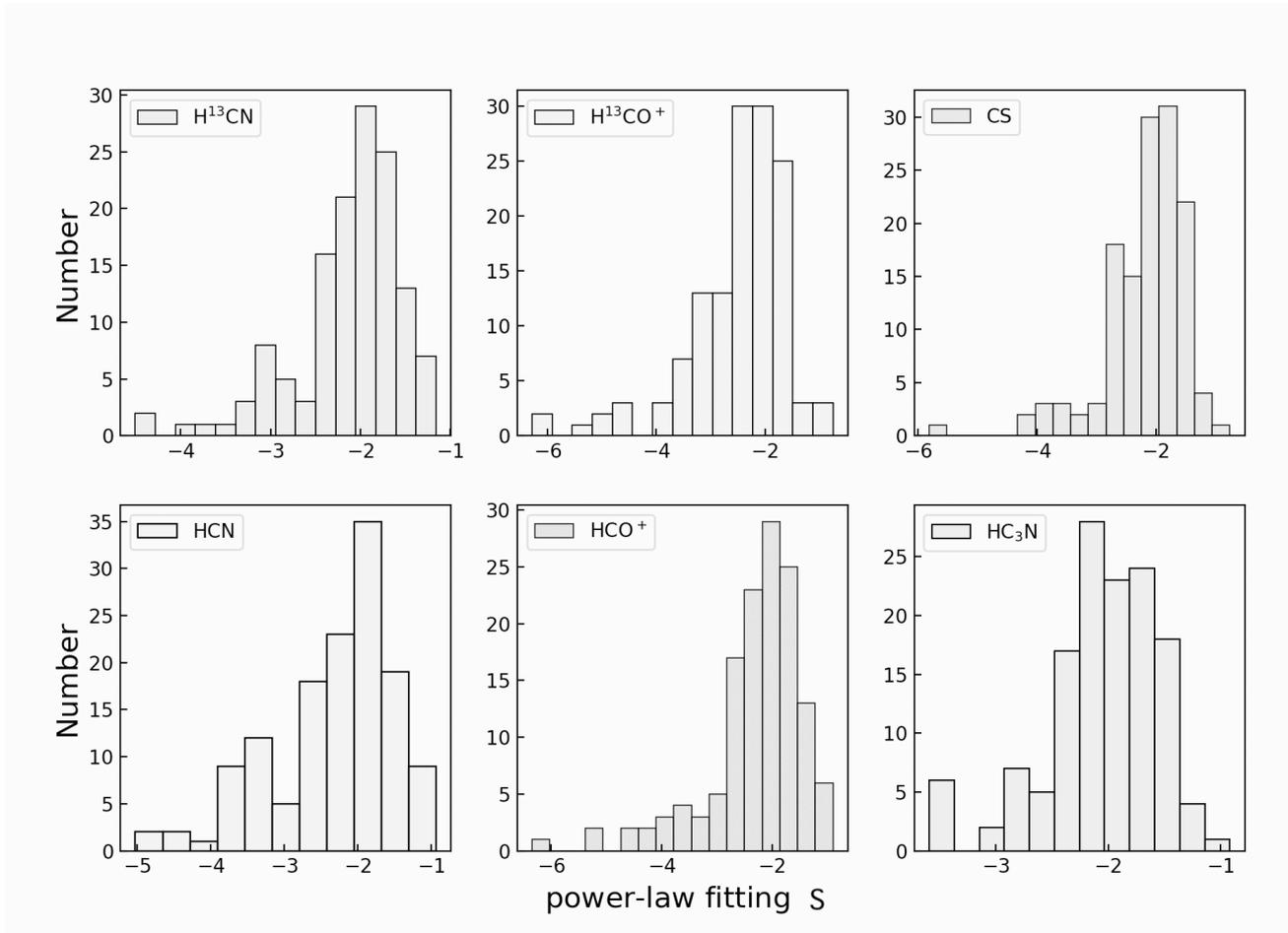

**Figure 3.** The distribution of the power-law tail slopes *s* for six tracers of all sources.

The brightness temperature is determined by taking the average value within a 30 arcsec circular region centered on the source. By calculating the ratio of the brightness temperatures of HCO$^+$ to H$^{13}$CO$^+$ and applying Equations 2 and 3, we derived the optical depths of HCO$^+$ and H$^{13}$CO$^+$. The upper panels of Figure 4 present the distributions of optical depth for the H$^{13}$CO$^+$ and HCO$^+$. The optical depth for the majority of sources of H$^{13}$CO$^+$ is approximately 0.2, whereas for HCO$^+$ it is around 10. It is evident that the optically thick assumption, $\tau_{\nu,12} > 1$, is always satisfied for HCO$^+$, consistenting with the findings presented in Xu et al. (2023). Employing the same approach, we have also determined the optical depth distribution for both HCN and H$^{13}$CN, as shown in the lower panels of Figure 4. Consistent with the findings for HCO$^+$ and H$^{13}$CO$^+$, HCN is much more opticlaly thick than H$^{13}$CN.

We note that both HCN and HCO$^+$ are also susceptible to the effects of missing flux, although the influence on the power-law component may be somewhat mitigated. Even though the extended or diffuse emission of HCN and HCO$^+$ has been filtered out in our ALMA data, it seems that their emission is still not very sensitive to dense gas structures or gravitationally bound dense gas, as indicated by the sublinearity in the scaling relations L$_{bol}$-to-L$_{mol}$ (p). On the other hand, the isotopic molecular lines H$^{13}$CO$^+$ and H$^{13}$CN suffer similar missing flux as their main lines but are much more optically thin, suggesting that the large optical depths are likely the predom-inant factor causing the sub-linear trends in the scaling relations L$_{bol}$-to-L$_{mol}$ for HCO$^+$ and HCN.

In summary, the J=1-0 transitions of HCN and HCO$^+$ are optically thick. When considering the density range of the power-law tail, which corresponds to regions where gravity is the predominant force, their high optical depths prevent them being the most suitable tracers for gas that is gravitationally bound.

Among the six tracers, H$^{13}$CN and HC$_3$N show the highest linearity in the scaling relations between L$_{bol}$ and L$_{mol}$ (p), indicating that H$^{13}$CN and HC$_3$N are very good tracers for gravitationally bound and dense star forming gas. These findings align with the perspective by Liu et al. (2020a), who argue that H$^{13}$CN and HC$_3$N emission can well trace the dense cores within massive clumps.

## 4 DISCUSSION

In this part, we mainly explore whether the power-law slope will change at different stages of evolution, and the correlation between the power-law slope and the star formation efficiency (SFE).

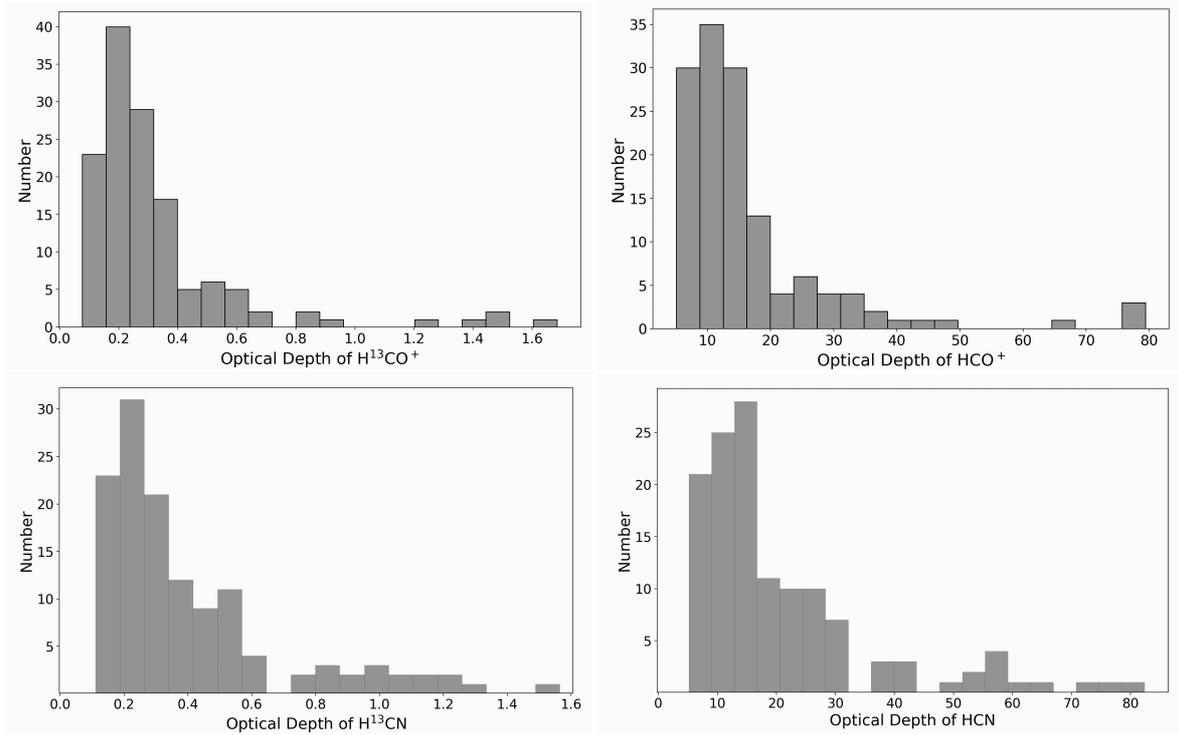

**Figure 4.** TopThe distribution of the optical depth for H$^{13}$CO$^+$ is depicted in the left panel, while the right panel illustrates the optical depth distribution for HCO$^+$. Bottom: The distribution of the optical depth for H$^{13}$CN is depicted in the left panel, while the right panel illustrates the optical depth distribution for HCN.

### 4.1 power-law slope *s*

As massive protoclusters evolve, the strong stellar feedback including outflow and ionization keep heating the gas, so the luminosity-to-mass ratio (L/M) of their natal clumps will increase. Therefore, L/M is often used for the evolutionary classification of dense star-forming clumps (Saraceno et al. 1996; Molinari et al. 2008; Stephens et al. 2016; Urquhart et al. 2018; Molinari et al. 2019; Urquhart et al. 2022; Xu et al. 2024b,a). The mass estimates for our sample sources are taken from Liu et al. (2020a). Figure 6 presents the relationship between the power-law slope *s* and the luminosity-to-mass (L/M) ratio within three separate distance ranges: less than 3 kpc, between 3 and 6 kpc, and beyond 6 kpc. Within any of these distance intervals, there is no clear trend between the *s* and the L/M ratio. The possible reason is that within this range of L/M ratio, these massive clumps are relatively stable and exhibit similar I-PDF distribution patterns, indicating that they have comparable forms of intensity distribution. Consequently, there is no discernible trend between the *s* and the L/M ratio.

We additionally present an analysis depicting the relationship between the power-law slope *s* and the surface density of clumps within three discrete distance ranges. The surface density is calculated by dividing the mass by the surface area. The volume area is $\pi R^2$, where R (compiled by Liu et al. 2020a) is the radius of the clump. Figure 7 also shows that there is no clear trend between the *s* and the clump density within three separate distance ranges. This figure illustrates that within this range of surface density, both high and low surface density massive clumps exhibit similar I-PDF distribution patterns, indicating that they have comparable forms of intensity distribution.

L$_{bol}$-to-L$_{mol}$ ratios are often interpreted in terms of the SFE (Evans et al. 2020b). Understanding the regulatory factors of SFE within star-forming regions remains a fundamental question in star formation research. Recent studies suggest that the SFE remains relatively stable above the critical density threshold, supporting a scenario in which the SFE in dense gas is approximately constant, albeit with considerable scatter (e.g., Mattern et al. 2024, Jiao et al. submitted). Here, we further explore whether SFE is influenced by gas accumulation, which can be indicated by the power-law tails from I-PDFs.

A flatter power-law tail, suggests a greater accumulation of high-density gas. Figure 8 presents a correlation plot between the L$_{bol}$-to-L$_{mol}$($p$) ratios and the slope *s*. Observations from this figure indicate that SFE, as traced by the L$_{bol}$-to-L$_{mol}$ ratios, decreases with the power-law slope *s*. We employ the Spearman rank correlation (SRC) test to assess the significance of the correlations in Figure 8. The SRC test, being based purely on ranks instead of the original quantities, is robust against outliers. We derive SRC coefficients ($\rho$) for the pairs of parameters, where $\rho$ ranges within ±1. The more positive/negative $\rho$ is, the more positively/negatively correlated the ranks. $\rho$ = 1 suggests a strong correlation, $\rho$ = -1 a strong anti-correlation, and $\rho$ = 0 no correlation. The SRC coefficients for H$^{13}$CN, H$^{13}$CO$^+$, CS, HCN, HCO$^+$, HC$_3$N are -0.35, -0.27, -0.42, -0.52, -0.43 and -0.37, respectively. The SRC test supports a weak anticorrelation between the dividing and the power law slope *s*.

The slope of L$_{bol}$-to-L$'_{mol}$ ratios and the slope *s* correlation for H$^{13}$CN, H$^{13}$CO$^+$, CS, HCN, HCO$^+$, HC$_3$N are -0.33 ± 0.07, -0.17 ± 0.05, -0.37 ± 0.06, -0.43 ± 0.05, -0.32 ± 0.05, -0.4 ± 0.08, respectively. We utilize the coefficient of determination, denoted as $\mathfrak{R}^2$ to assess the goodness of fit. The range of the $\mathfrak{R}^2$ value is from 0 to 1. The closer the value is to 1, the better the model fits the data. The $\mathfrak{R}^2$ values for H$^{13}$CN, H$^{13}$CO$^+$, CS, HCN, HCO$^+$, HC$_3$N are 0.16, 0.10, 0.24, 0.34, 0.23, and 0.15, respectively. The parameters indicate that a linear decline does not effectively describe the relationship between the slope $s$ and SFE, although there is indeed a slight decrease in SFE as the slope increases. This suggests that an increased aggregation of high-density gas might not necessarily enhance the star formation efficiency. Moreover, the gravitationally bound dense gas may be more consumed or dispersed by stellar feedback in more evolved sources, which may contribute to the observed scatter in SFE.

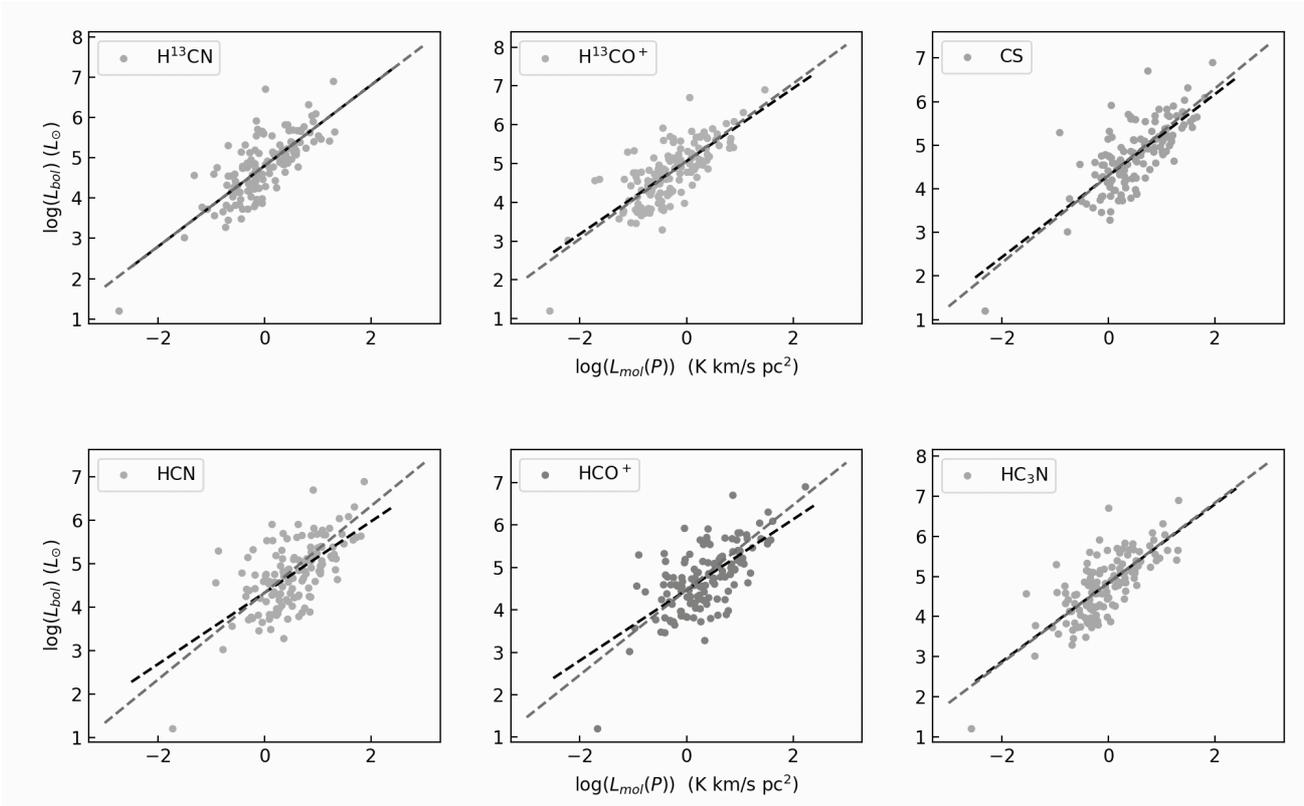

**Figure 5.** The $L_{bol}$-$L_{mol}$(p) scaling relations of six tracers. The fit towards the six different tracers is shown in black dash line. The green dash line is one-by-one line.

### 4.2 power-law slope $s$ of $H^{13}CO^+$

Schneider et al. (2016) proposed that in the column density probability distribution function (N-PDF), the slope $s'$ of the power-law tail can be translated into the exponent of an equivalent density distribution $\rho \propto r^{-\alpha}$, where r represents the radius. Given that our distribution is I-PDF, to convert the power-law tail slope $s$ into the exponent $\alpha$, we should assume that the tracer is optically thin, which allows the tracer's intensity to be directly proportional to the column density, thus fulfilling the conversion relationship previously mentioned.

Based on our previous work, $H^{13}CO^+$ can be taken as optically thin line even towards the most massive cores (e.g. Liu et al. 2020b; Xu et al. 2023; Liu et al. 2022). For a spherical geometry that represents either a single core or a collection of cores, we apply the conversion formula proposed by Schneider et al. (2016), which is $\alpha_c$=1-2/s. In the simplified picture of the free-fall of a collapsing sphere, $\alpha_c$ = 2 for early stages and $\alpha_c$ = 1.5 after a singularity formed at the centre of the sphere (Shu 1977; Larson 1969; Penston 1969; Whitworth &

Summers 1985). The values we obtain for $\alpha_c$ vary around 1.4 to 3 which is consistent with the spherical free-fall scenario.

Figure 9 illustrates the ratio of $L_{bol}$-to-$L_{mol}(p)$ plotted against $\alpha_c$. The majority of the data points fall within the range of $\alpha_c$ from 1.3 to 2.4. There are only three points with $\alpha_c$ values exceeding 2.5, which are distinctly separated from the distribution of the other points. Therefore, these three points are treated as outliers and are excluded from the fitting process. The correlation between $\alpha_c$ and the $L_{bol}$-to-$L_{mol}(p)$ ratio is weak, with a slope of -0.67 and an $\mathfrak{R}^2$ value of 0.11. This suggests that the star formation efficiency (SFE), as indicated by the $L_{bol}$-to-$L_{mol}(p)$ ratio, shows a slight tendency to decrease with increasing $\alpha_c$. In conjunction with Figure 8, we argue that a higher slope $s$ or a higher $\alpha_c$, signifing a more substantial aggregation of gas, does not correspond to a higher SFE. In addition, when $\alpha_c$ falls within the interval from 1.8 to 2.25, the SRC coefficient is -0.07, which is approximately zero. It means there is no discernible trend of variation between the $L_{bol}$-to-$L_{mol}(p)$ ratio and $\alpha_c$ when $\alpha_c$ is larger than 1.8, indicating that those likely collapsing clumps also do not show obvious changes in SFE.

However, the purely spherical free-fall picture appears unlikely to apply to the gas which defines the power-law distributions within PDFs, as the gas observed in the ATOMS survey exhibits hub-filament systems (Zhou et al. 2022). For the singular polytropic cylinders, which depict filaments, we use $\alpha_f$=1-1/s (Schneider et al. 2016). The values we obtain for $\alpha_f$ vary around 1.2 to 1.8 which are consistent with self-gravitating but non-collapsing filament models (Toci & Galli 2015; Myers 2015). In this case, filamentary gas that has a sufficient mass per length ratio is centrally concentrated and self-gravitating, but does not need to collapse globally (Schneider et al. 2016).

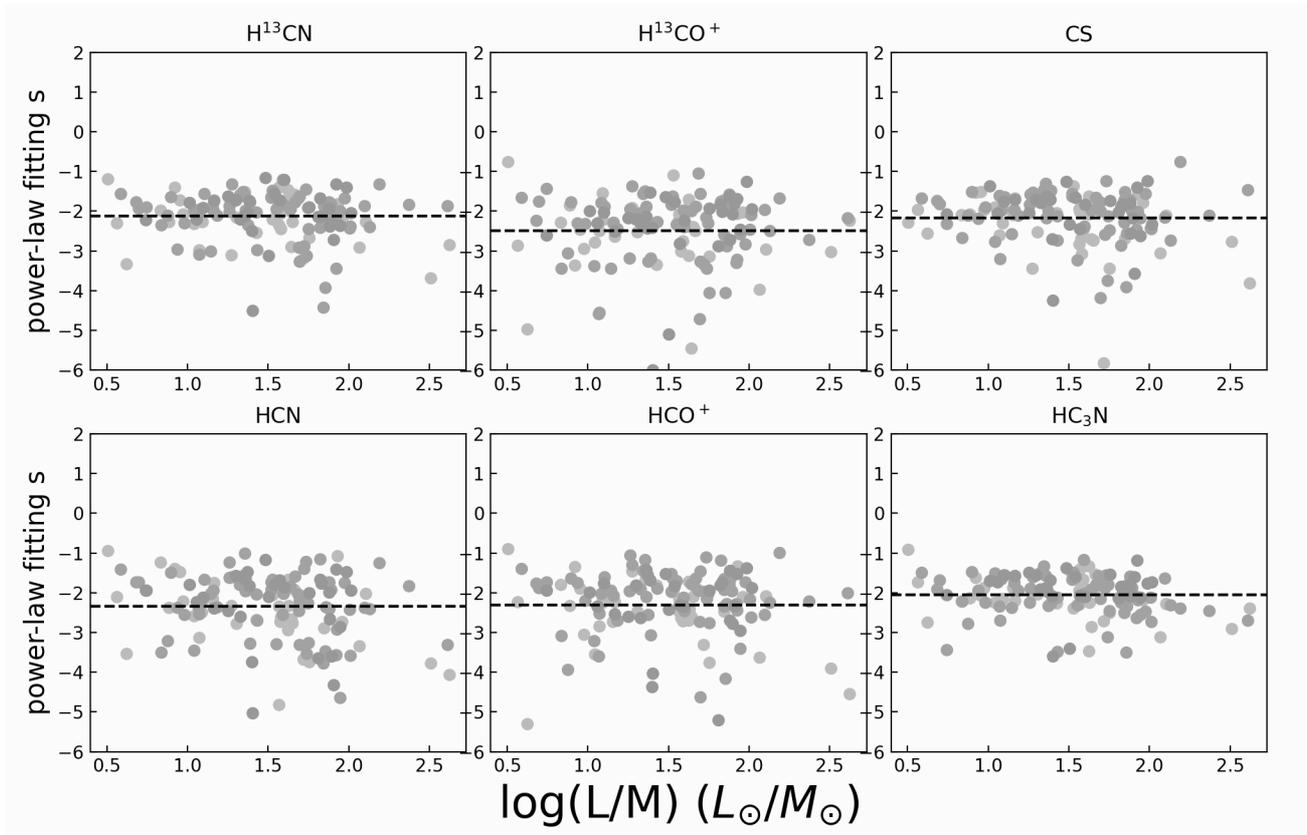

**Figure 6.** The relationship between the power-law slope *s* and the luminosity-to-mass (L/M) ratio within three separate distance ranges. The green, brown and purple dots repersent the distance less than 3 kpc, between 3 to 6 kpc and lager than 6 kpc, respectively. The black dash line shows the median value of power-law fitting.

## 5 CONCLUSION

We report the detection of emissions of different molecules HCN, $HCO^+$, $H^{13}CO^+$, and $H^{13}CN$ J=1-0, CS J=2-1, $HC_3N$ J=11-10 towards 135 sources. The intensity PDF for these molecular tracers are presented, incorporating both log-normal fits and power-law tail fits. All traces indicate a log-normal distribution in the PDFs and the presence of power-law tails. We have determined the molecular line luminosities for the power-law tail component, denoted as $L_{mol}(p)$, of the distribution function. We arrived at the following conclusions:

(i) The slope of the $L_{bol}$-to-$L_{mol}$(p) scaling relations for $H^{13}CN$, $H^{13}CO^+$, CS, HCN, $HCO^+$, $HC_3N$ are $1.0 \pm 0.07$, $0.94 \pm 0.07$, $0.93 \pm 0.07$, $0.82 \pm 0.09$, $0.83 \pm 0.08$, $0.98 \pm 0.08$, respectively. These indicate a closer association between the gas in the power-law tail and star formation processes. $H^{13}CN$ and $HC_3N$, showing highest linearity in the $L_{bol}$-to-$L_{mol}$(p) scaling relations, may serve as the most effective tracers for the dense molecular gas mass among the six molecular line tracers. In contrast, HCN and $HCO^+$ exhibit a significantly sublinear trend in the $L_{bol}$-to-$L_{mol}$(p) scaling relations due to the effects of optical thickness, which diminishes their effectiveness as tracers for gravitationally bound gas.

(ii) Across three distinct distance intervals — those less than 3

kpc, those ranging from 3 to 6 kpc, and those greater than 6 kpc — there is no discernible trend between the slope *s* of the power-law tails and the L/M ratio or the density of the clumps.

(iii) Correlations between the ratios of $L_{bol}$ to $L_{mol}(p)$, which can be interpreted as SFE, and the power-law tail are explored and a weak anticorrelation is found. This suggests that molecular clouds with significant gas accumulation may still exhibit comparatively low star formation efficiencies.

(iv) For $H^{13}CO^+$, the power-law fitting can be converted into the exponent $\alpha$ of an equivalent density distribution. Our observations indicate that the star formation efficiency (SFE) is very weakly anticorrelated with $\alpha$, indicating that SFE does not change remarkably as a clump collapses.

## ACKNOWLEDGEMENTS

This work has been supported by the National Key R&D Program of China (No. 2022YFA1603100). Z.C. acknowledges support from the National Natural Science Foundation of China (NSFC), through grants No. 12403028, the Basic Research Program of Shanxi Provence (202403021222272). T.L. acknowledges support from the National Natural Science Foundation of China (NSFC), through grants No. 12073061 and No. 12122307, the Tianchi Talent Program of Xinjiang Uygur Autonomous Region, and the international partnership program of the Chinese Academy of Sciences, through grant No. 114231KYSB20200009. SRD acknowledges support from the Fondecyt Postdoctoral fellowship (project code 3220162) and ANID BASAL project FB210003. H.-L. Liu is supported by Yunnan Fundamental Research Project (grant No. 202301AT070118, 202401AS070121), and by Yunnan Xingdian Talent Support Plan–Youth Project. DL is a new Cornerstone investigator. C.W.L. is supported by the Basic Science Research Program through the NRF funded by the Ministry of Education, Science and Technology (NRF-2019R1A2C1010851) and by the Korea Astronomy and Space Science Institute grant funded by the Korea government (MSIT; project No. 2024-1-841-00). SRD acknowledges support from the Fondecyt Postdoctoral fellowship (project code 3220162) and ANID BASAL project FB210003. L.B. gratefully acknowledges support by the ANID BASAL project FB210003. MJ acknowledges the support of the Research Council of Finland Grant No. 348342. K.T. was supported by JSPS KAKENHI (Grant Number JP20H05645). PS was partially supported by a Grant-in-Aid for Scientific Research (KAKENHI Number JP22H01271 and JP23H01221) of JSPS.

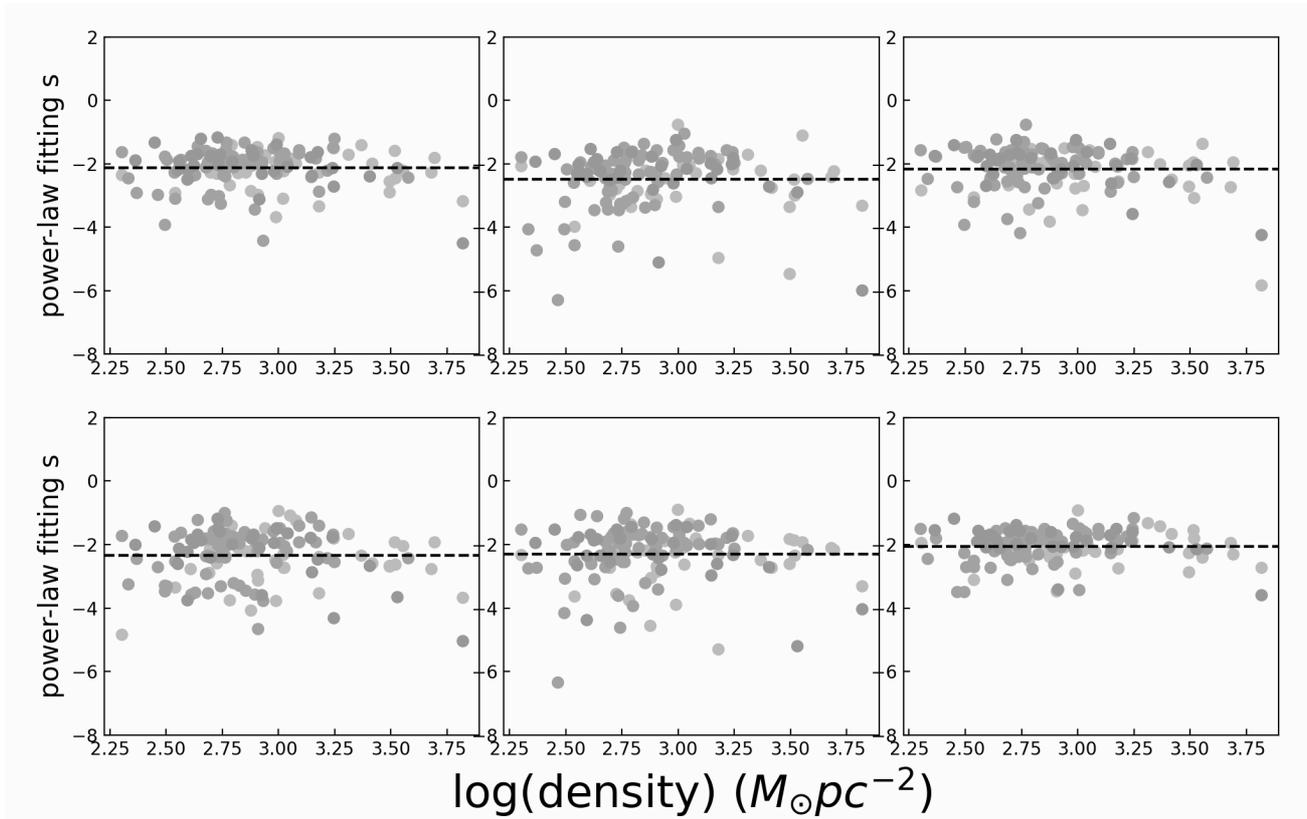

**Figure 7.** The relationship between the power-law slope *s* and the clump density within three separate distance ranges. The green, brown and purple dots represent the distance less than 3 kpc, between 3 to 6 kpc and lager than 6 kpc, respectively. The black dash line show the median value of power-law fitting.

This paper makes use of the following ALMA data: ADS/JAO.ALMA#2019.1.00685.S. ALMA is a partnership of ESO (representing its member states), NSF (USA), and NINS (Japan), together with NRC (Canada), MOST and ASIAA (Taiwan), and KASI (Republic of Korea), in cooperation with the Republic of Chile. The Joint ALMA Observatory is operated by ESO, AUI/NRAO, and NAOJ. D.L. is a New Cornerstone Investigator. G.G. and L.B. gratefully acknowledge support by the ANID BASAL project FB210003.

## DATA AVAILABILITY

The raw data are available in ALMA archive. The derived data underlying this article are available in the article and in its online supplementary material.

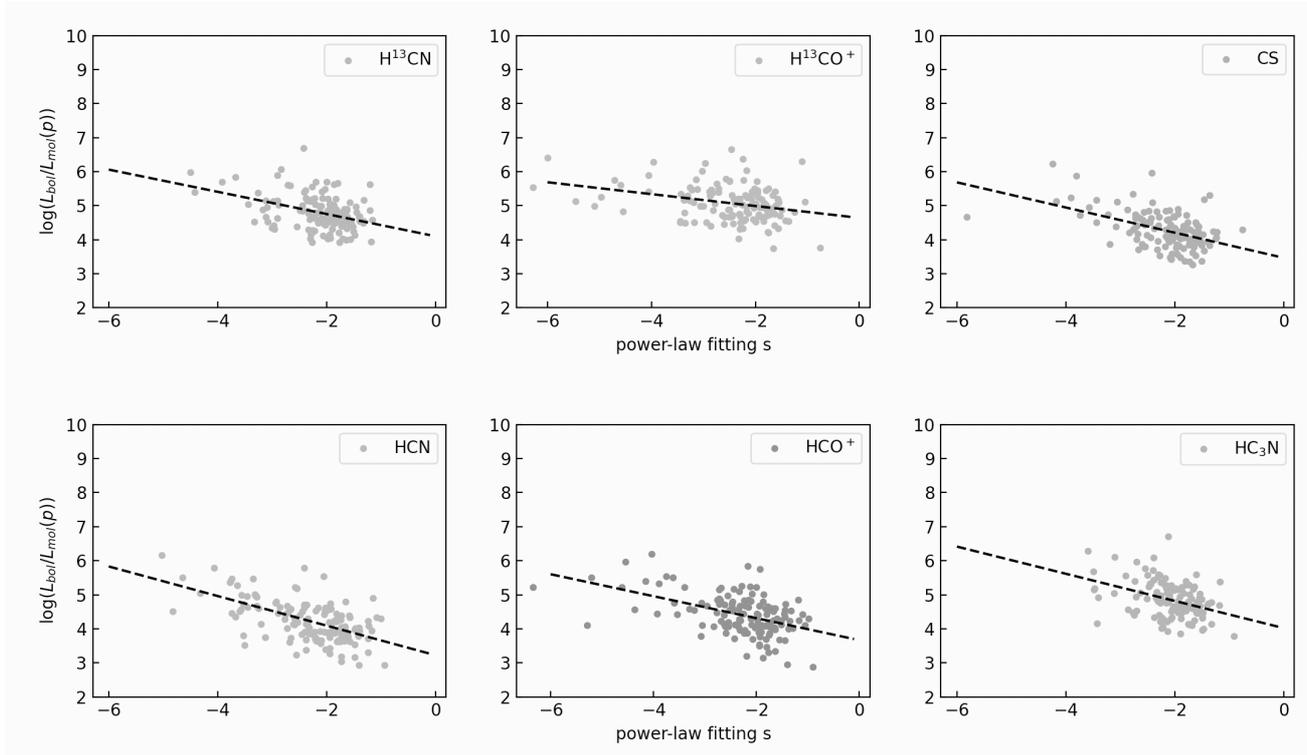

**Figure 8.** $L_{bol}$-to-$L_{mol}(p)$ versus slope $s$ of power-law fitting. The black dash line show the linear fitting.

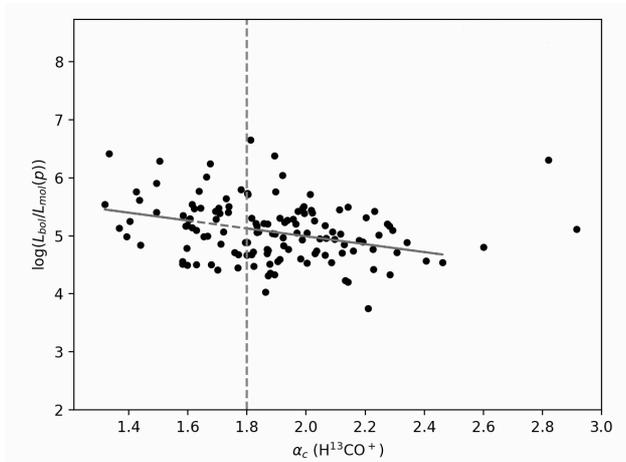

**Figure 9.** $L_{bol}$-to-$L_{mol}(p)$ versus $\alpha_c$. The red bashed line is the $\alpha_c$ equal to 1.8. The green dashed line shows the linear fitting.

This paper has been typeset from a TeX/LaTeX file prepared by the author.